\documentclass[twocolumn,showpacs,preprintnumbers,amsmath,amssymb]{revtex4}
\usepackage{latexsym} 
\usepackage{youngtab}
\usepackage{graphicx}
\def\qed{\hbox{${\vcenter{\vbox{                          
 \hrule height 0.4pt\hbox{\vrule width 0.4pt height 6pt
   \kern5pt\vrule width 0.4pt}\hrule height 0.4pt}}}$}}
\def\shiftleft#1{#1\llap{#1\hskip 0.04em}}
\def\shiftdown#1{#1\llap{\lower.04ex\hbox{#1}}}
\def\thick#1{\shiftdown{\shiftleft{#1}}}
\def\b#1{\thick{\hbox{$#1$}}}
\newcommand{\bea}{\begin{eqnarray}}
\newcommand{\eea}{\end{eqnarray}}
\newcommand{\be}{\begin{equation}}
\newcommand{\ee}{\end{equation}}
\begin{document}
\title{Metaspin and dirishonic dark matter}
\renewcommand{\thefootnote}{\fnsymbol{footnote}}
\author{Alfons J. Buchmann 
and Michael L. Schmid}
\affiliation{Institut f\"ur Theoretische Physik, Universit\"at T\"ubingen,
Auf der Morgenstelle 14, D-72076 T\"ubingen, Germany}
\email{alfons.buchmann@uni-tuebingen.de}
\email{micha.l.schmid@gmx.net}

\begin{abstract}
The antisymmetry requirement of rishon bound state wave functions 
suggests a new rishon quantum number called $M$ spin.
From $M$ spin conservation and the Nussinov-Weingarten-Witten theorem 
we predict the existence of a stable pseudoscalar dirishonic meson $\zeta$ 
that is lighter than the lightest neutrino. Its mass 
is estimated as $m_{\zeta} = 10^{-9}$ eV. This particle could make up  
the major part of cold dark matter in the Universe.   
\end{abstract}

\smallskip

\maketitle

\section{Introduction}
\label{sec:intro}
Among the various approaches to lepton and quark substructure~\cite{sou92}, 
the rishon model, which has only two massless spin 1/2 
building blocks $T$ and $V$ interacting via color 
and hypercolor forces, stands out because of its simplicity.
The rishon model~\cite{har79,har81,har82,har83,har84} provides answers 
to fundamental questions that remain unanswered within the standard model.
For example, it explains why color triplet quarks 
have fractional whereas color singlet leptons have integer charges with 
$\vert Q \vert \le 1$, and why the electroweak anomalies of leptons
and quarks cancel~\cite{har84}. Furthermore, it explains the origin of weak 
interactions as a residual force on the rishon bound state 
level~\cite{har84,abms09}, and suggests an explanation of 
the generation number~\cite{abms05}. 

On the other hand, a major problem has not been sufficiently clarified. 
This concerns the requirement that preon bound state wave functions 
must be antisymmetric under the interchange of two spin 1/2 preons 
(Pauli exclusion principle). 
Originally, it was proposed that the overall antisymmetry of rishon bound 
states lies in the space-time degrees of freedom~\cite{har81}, 
specifically, that three massless rishons in the fundamental $(1/2,0 +0,1/2)$ 
representation of the relativistic space-time group $SL(2,C)$ combine 
to a totally antisymmetric product state.   
Another approach employs quaternions, a noncommutative 
generalization of complex numbers connected with four spatial dimensions, 
and quaternionic quantum mechanics~\cite{adl94}.

The Pauli exclusion principle is a general quantum mechanical concept 
based on the indistinguishability of identical particles 
and the permutation group ${\cal S}_n$, and the problem of 
constructing fully antisymmetric rishon bound states may be
formulated in a still different manner. 
The problem can be most clearly seen in the case 
of the color and hypercolor singlet positron $e^{+}(TTT)$, 
which consists of three $T$ rishons. 
This is studied in more detail in Sec.~\ref{sec:pauli},
where we show that the Pauli principle can be satisfied
if rishons carry an additional conserved SU(2) like quantum number, 
called metaspin $M$. The possible origin of metaspin is discussed
in Sec.~\ref{sec:meta}.

As an interesting consequence of metaspin conservation we find 
in combination with the Nussinov-Weingarten-Witten (NWW)  
theorem~\cite{nww83} an argument for the existence of a stable
dirishonic scalar bound state, which could be a  
candidate for cold dark matter. The argument goes like this.
The NWW theorem applied to rishons requires a dirishonic scalar bound state 
that is lighter than the lightest three-rishon fermionic bound state, 
i.e. the neutrino. We call this particle scalarino $\zeta$.  
\begin{figure}
\resizebox{0.45\textwidth}{!}{\includegraphics{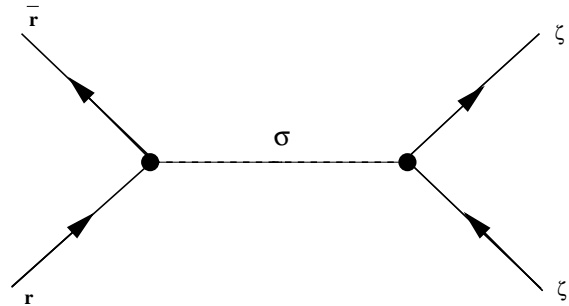}}
\caption{\label{figure:scalar_interaction} 
Formation of a heavy scalaron 
$\sigma$ with mass $m_{\sigma} > m_G = 10^{16}$ GeV from a rishon-antirishon 
pair $r{\bar r}$ and its decay into scalarinos $\zeta$.}
\end{figure}
As a color and hypercolor singlet 
with spin $S=0$ and electric charge $Q=0$, the scalarino does not interact 
with ordinary matter via electromagnetic, weak or strong forces 
but only via short-range scalar interactions and gravitation. 
Scalarinos may have been produced via primordial rishon-antirishon 
annihilation into a heavy scalar boson, called scalaron $\sigma$, 
which subsequently decays into two scalarinos 
as depicted in Fig.~\ref{figure:scalar_interaction}. 
In Sec.~\ref{sec:nww} we argue that the scalaron must have 
a mass above the grand unified theory (GUT) scale of $10^{16}$ GeV
to prevent fermionic rishon bound states from having long-range scalarino
mediated interactions, which are not observed.
Furthermore, we will see that scalarinos have metaspin $M=1$.  
Therefore, they cannot decay into massless gauge bosons with metaspin 
$M=0$. This explains their stability once they have been formed. 
Possible consequences for cosmology and particle physics are studied 
in Sec.~\ref{sec:scalar} and Sec.~\ref{sec:dark}. 

\section{Rishon bound states, Pauli principle, and metaspin}
\label{sec:pauli}

\subsection{Statement of the problem}

Rishons are spin 1/2 particles. Therefore, according to the 
Pauli exclusion principle, rishon bound states 
must be fully antisymmetric under the interchange of two 
rishons of the same type. The consequences of this requirement 
are most transparent for the positron $e^{+}$, 
the rishon content of which is $TTT$. 
The positron is color and hypercolor neutral. As a result, 
its color-hypercolor product state is 
necessarily fully symmetric with respect to permutation of the rishon 
constituents. 

We assume that the $e^+$ orbital ground state 
is totally symmetric under permutation.
Because the spin wave function of the $e^+$ 
is mixed symmetric, there must be another 
degree of freedom in which the three $T$ 
rishons are in a state of mixed symmetry 
to generate an overall totally antisymmetric wave function 
for the positron. 
This additional degree of freedom 
cannot be SU(2) isospin as in the quark model. For $T$ and $V$ rishons
there is no global SU(2) isospin symmetry because they
are in different color representations~\cite{har81} as is evident 
from Table~\ref{tab:preons}. 
Thus, we postulate that rishons carry a new SU(2) quantum number, 
called metaspin $M$. 
Although the above argument requires a noncovariant separation of spin 
and space coordinates, it is not in conflict with special 
relativity. For particles with nonzero rest mass, it is possible to define 
orbital and spin symmetries separately~\cite{eli90}.

\begin{table}[htb]
\begin{tabular}{l c c c c c c c c}
\hline
Rishon & H & C & Q & ${\cal P}$ & $ \Upsilon $ & $\Pi$ & S & M  \\
\hline
$T$ & $3$ & $3$ & $ +\frac{1}{3} $ & $+\frac{1}{3}$ & $+\frac{1}{3}$ 
& -1 & $\frac{1}{2}$ &  $\frac{1}{2}$ \\
\hline
$V$ & $3$ & $\bar{3}$ & 0 & $+\frac{1}{3}$ & $-\frac{1}{3}$ & +1 & 
$\frac{1}{2}$ &  $\frac{1}{2}$  \\    
\hline
$\bar{V}$ & $\bar{3}$ & $3$ & 0 & $-\frac{1}{3}$ & $+\frac{1}{3}$ & -1 
& $\frac{1}{2}$ &  $\frac{1}{2}$  \\
\hline
$\bar{T}$ & $\bar{3}$ & $\bar{3}$ & $ -\frac{1}{3} $  & $-\frac{1}{3}$ & 
$-\frac{1}{3}$ & +1 & $\frac{1}{2}$ &  $\frac{1}{2}$ \\
\hline
\end{tabular}
\caption{The hypercolor (H), color (C), electric charge (Q), rishon number
$({\cal P})$, $\Upsilon$ number $(\Upsilon)$, 
intrinsic parity ($\Pi$), spin (S), and 
metaspin (M) of rishons and antirishons.
The quantum numbers ${\cal P}$ and $\Upsilon$ are defined 
as ${\cal P}=(n(T) + n(V))/3$ and
$\Upsilon=(n(T)-n(V))/3$, where $n(T)$ and $n(V)$ are the number of $T$ and
$V$ rishons, and $Q= ({\cal P}+\Upsilon)/2$. } 
\label{tab:preons}
\end{table}

The Pauli principle demands that the product 
wave function of rishon bound states $\vert B \rangle$ 
\begin{equation}
\label{pwv}
\left| B \right.\rangle =\left| O \right.\rangle 
\times \left| H \right.\rangle \times 
\left| C \right.\rangle 
\times \left| S \right.\rangle \times \left| M \right.\rangle \mbox{,}
\end{equation}
must be a fully antisymmetric representation of the direct product group 
of orbital (O), hypercolor (H), color (C), spin (S) and metaspin (M) groups
\begin{equation}
O(3)_{O} \times  SU(3)_{H} \times SU(3)_{C} \times SU(2)_{S} 
\times SU(2)_{M} \mbox{.} 
\end{equation}
Individual $T$ and $V$ rishons 
are represented by Young diagrams according to the quantum number 
assignments of Table~\ref{tab:preons} as
\be
\Yvcentermath1
\vert T \rangle  =   \yng(1)_{\ O} \times \yng(1)_{\ H} 
\times \yng(1)_{\ C} \times \yng(1)_{\ S} \times \yng(1)_{\ M}  
\ee 
\be
\Yvcentermath1
\vert V \rangle  =  \yng(1)_{\ O} \times \yng(1)_{\ H} 
\times \yng(1,1)_{\ C} \times \yng(1)_{\ S} \times \yng(1)_{\ M}  
\ee
noting that the $V$ rishon is an SU(3)$_C$ antitriplet ${\bf 3}_C$ 
(see Table~\ref{tab:preons}) represented by two boxes just as 
an antisymmetric two-rishon state.  

We then form outer products of multi-rishon states in 
orbital, color, hypercolor, spin, and metaspin space separately. 
In the following, we discuss these outer products for 
fermions (leptons, quarks, and hyperquarks~\cite{comment1}) 
and spin 1 bosons (dirishons and six-rishons). 
Dirishonic bound states with spin 0 are discussed in Sec.~\ref{sec:nww}
and Sec.~\ref{sec:scalar}.

\begin{table*}[htb]
\begin{center}
\begin{tabular}{ l c c c c c c c c c c }
\hline
& Rishon content & Bound state & H & C & ${\cal{P}}$ & 
$\Upsilon$ & $Q$ & S & M  \\ 
\hline
& $\left( VVV \right)$ & $\left( \nu_{e}, \nu_{\mu}, \nu_{\tau} \right)$
& 1 & 1 & $+1$ &$ -1$ &$ 0$ & $\frac{1}{2}$ &  $\frac{1}{2}$   \\
Leptons & & & & & & & \\
& $\left(\bar{T}\bar{T}\bar{T} \right)$ & $\left( e^{-}, \mu^{-}, \tau^{-} 
\right)$ & 1 & 1 & $ -1$ & $-1$ & $-1$ & $\frac{1}{2}$ &  $\frac{1}{2}$  \\
\hline
& $\left( TTV \right)$ & $\left( u, c, t \right)$ & 1 & 3 & +1 
& $+\frac{1}{3}$ & $+\frac{2}{3}$ & $\frac{1}{2}$ &  $\frac{1}{2}$   \\
Quarks & & & & & & & \\
& $\left(\bar{T}\bar{V}\bar{V} \right)$ & $\left( d, s, b \right)$ & 1 & 
3 & -1 & $+\frac{1}{3}$ & $-\frac{1}{3}$ & $\frac{1}{2}$ &  $\frac{1}{2}$  \\ 
\hline
& $\left( TT\bar{V} \right)$ & $\left( \tilde{u},\tilde{c},\tilde{t} \right)$ 
& 3 & 1 & $+\frac{1}{3}$ & +1 & $+\frac{2}{3}$ & $\frac{1}{2}$ &  
$\frac{1}{2}$  \\  
Hyperquarks & & & & & & & \\
& $\left( \bar{T}VV \right)$ & $\left( \tilde{d}, \tilde{s}, \tilde{b}
\right)$ & 3 & 1 & $+\frac{1}{3}$ & -1 & $-\frac{1}{3}$ & 
$\frac{1}{2}$ &  $\frac{1}{2}$   \\
\hline
\end{tabular}  
\caption{Allowed three-rishon bound states representing leptons, quarks and
hyperquarks and their quantum numbers. Formally, the
hyperquarks are obtained from the corresponding quarks by interchanging: 
$V \leftrightarrow \bar{V}$.}
\label{tab:fermions}
\end{center}
\end{table*}
\subsection{Leptons, quarks, and hyperquarks}
We begin with the orbital wave function. 
For the three-rishon bound states listed 
in Table~\ref{tab:fermions} and their antiparticles,  
we assume that the O(3)$_{O}$  orbital wave function is the fully 
symmetric Young tableau $\yng(3)_{\ O}$. 

Next, we construct the outer product of three rishons in hypercolor and
color spaces separately, and in a second step, 
form hypercolor-color inner product states.
For the hypercolor and color singlet leptons containing three preons 
of the same type (see Table~\ref{tab:fermions}) one obtains, e.g., for 
the positron $e^+(TTT)$
\begin{equation}
\label{youngcolorhypercolor}
\Yvcentermath1
\yng(1,1,1)_{\ H} \times \yng(1,1,1)_{\ C} =
\yng(3)_{{\ e^+}_{_{(HC)}}}.  
\end{equation}
Analogous results are obtained for the neutrino, so that leptons 
have fully symmetric hypercolor-color product states.

For quarks, the hypercolor product state is again the fully antisymmetric 
$SU(3)_H$ singlet, whereas in color space a  
color triplet is required. For example, for a $u(TTV)$ quark, 
the outer product (denoted by $\otimes$) of the two 
color-triplet $T$ and the color-antitriplet $V$ rishons leads to 
the following irreducible $SU(3)_C$ representations 
with color dimensions as subscripts:
\begin{widetext} 
\bea
\label{youngquark}
\Yvcentermath1
\left (\yng(1)_{\ {3}} \otimes \yng(1)_{\ {3}} \right )
\Yvcentermath1 
\otimes \yng(1,1)_{\ {\bar {3}}} 
& = & \left (\Yvcentermath1 \yng(1,1)_{\ {3}_A} 
+ \yng(2)_{{\ {6}}_{S}} \right )
\Yvcentermath1 
\otimes \yng(1,1)_{\ {\bar 3}} \nonumber \\
& = & 
\Yvcentermath1
\left( \yng(1,1)_{{\ {3}}_{A}} \otimes  \yng(1,1)_{\ {\bar 3}} \right ) 
\ + \  
\left (\yng(2)_{{\ {6}}_{S}} \otimes  \yng(1,1)_{\ {\bar 3}} \right )
\nonumber \\
& = &  
\Yvcentermath1
\yng(2,2)_{\ {\bar 6}_{S}} + \  \yng(2,1,1)_{{3}_{A}} 
+ \ \yng(2,1,1)_{{3}_{}} + \ \yng(3,1)_{\ {15}_{}} \longrightarrow 
\yng(2,1,1)_{{3}_{A}}. 
\eea
\end{widetext}
The same irreducible representations are obtained if one first 
combines the color-triplet $T$ with the color-antitriplet $V$. 
Because the latter is distinguishable from the former, 
we need only antisymmetrize the $T$ rishons.
This eliminates the diagrams arising from the symmetric $(TT)_{6_S}$.
From the remaining two diagrams generated by the coupling of the 
antisymmetric $(TT)_{3_A}$ and $(V)_{\bar 3}$ Young schemes, 
only the antisymmetric color triplet $3_A$ (indicated by the arrow)
has the correct total color symmetry.
While Eq.(\ref{youngquark}) contains 
the proper $SU(3)_C$ dimension of the $u$-quark, 
with respect to the permutation group ${\cal S}_{3}$, 
we have only three rishons and the permutational symmetry 
in hypercolor-color space is represented by three boxes 
in a single row as on the right-hand side of Eq.(\ref{youngcolorhypercolor}).

For leptons and quarks, our results concerning the permutational symmetry 
of the combined hypercolor-color product states are consistent with 
those of Harari and Seiberg~\cite{har81}.  
For hyperquarks, e.g., ${\tilde u}(TT{\bar V})$ 
the role of color and hypercolor is interchanged, otherwise the 
construction of hypercolor-color
wave functions is analogous to that of quarks. 
Thus, for leptons, quarks, and hyperquarks, the inner product of 
hypercolor and color spaces is the completely symmetric 
representation conjugate to the fully antisymmetric spin-metaspin 
product representation constructed next.

From three rishons with spin 1/2 and metaspin 1/2 we first construct 
mixed symmetric outer product states in spin and metaspin space separately.
These mixed symmetric spin and metaspin parts are then 
combined to form a fully antisymmetric inner product wave function
common to all fermionic bound states. In principle, two mixed symmetric states
can combine to fully symmetric, mixed symmetric, 
and fully antisymmetric ${\cal S}_{3}$ product states
\bea
\label{SU4} 
\Yvcentermath1
\yng(2,1)_{\ S} \times \yng(2,1)_{\ M} & = & 
\Yvcentermath1
\yng(3)_{\ SM} +  2 \ \ \yng(2,1)_{\ SM}  \nonumber \\
& &   \Yvcentermath1
+ \ \ \yng(1,1,1)_{\ SM} \Yvcentermath1
\longrightarrow 
\yng(1,1,1)_{\ SM}. 
\eea
As indicated by the arrow we confine ourselves to the totally antisymmetric
product state. The reason for this choice is particularly evident 
for the positron, where the orbital-color-hypercolor product state  
totally symmetric (see above). 
As a result, the antisymmetry of fermionic rishon bound states lies in the 
combined spin-metaspin space.

The overall product states of fermionic rishon bound states are then
\begin{equation}
\Yvcentermath1
\yng(1,1,1)_{\ OHCSM} = \yng(3)_{\ O} \times \yng(3)_{\ HC} 
\times \yng(1,1,1)_{\ SM}, 
\end{equation}
i.e. totally 
antisymmetric with respect to permutations of the rishon constituents.
Thus, the metaspin hypothesis allows us to construct 
fully antisymmetric three-rishon bound states. 

In addition, metaspin explains why there are no leptons, quarks, 
and hyperquarks with spin $S=3/2$. 
Spin 3/2 states are totally symmetric, which requires 
fully antisymmetric  metaspin wave functions. 
However, this is impossible 
in SU(2)$_M$, where the maximum number of rows in any Young scheme 
is restricted to two. 
Interestingly, Harari and Seiberg~\cite{har81} could also exclude 
spin 3/2 lepton states due to the antisymmetry of their relativistic 
space-time wave function but needed an additional argument to exclude 
spin 3/2 quarks.
\subsection{Dirishons with spin ${\bf S=1}$}
The dirishons $N(VV)$, $U(T{\bar V})$, and $\tilde{U}(TV)$ 
with spin $S=1$ have been discussed in detail in Ref.~\cite{abms09}, 
where we have shown 
that the neutralons $N(VV)$ transform hyperquarks into quarks, while  
the $U$ and $\tilde{U}$ bosons are responsible for transitions between 
leptons and quarks and between leptons and hyperquarks respectively. 
Hence, it suffices to reproduce 
in Table~\ref{tab:dipreons} the quantum numbers of these SU(9) gauge bosons.
 
As in the case of the three-rishon bound states, the  
hypercolor-color product state of dirishons is fully symmetric.
\begin{table}[h]
\begin{center}
\begin{tabular}{l c c c c c}
\hline
Dirishon & H & C & ${\cal{P}}$ & $\Upsilon$ &  $Q$ \\ 
\hline
$N\left(VV\right)$ & $\bar{3}$ & $ 3 $ & $ +\frac{2}{3} $ & $ -\frac{2}{3} $ 
& 0 \\ 
\cline{1-6}
$\bar{N}\left(\bar{V}\bar{V}\right) $ & $3$ & $ \bar{3} $ & $ -\frac{2}{3} $ & 
$ +\frac{2}{3} $ & 0 \\   
\hline\hline
$U\left(T\bar{V}\right) $ & 1 & $\bar{3}$ & $ 0 $ & $ +\frac{2}{3} $ & $ 
+\frac{1}{3}$ \\ 
\cline{1-6}
$\bar{U}\left(\bar{T}V\right) $ & 1 & $3$ & $ 0 $ & $ -\frac{2}{3} $ 
& $ -\frac{1}{3} $ \\ 
\hline
$\tilde{U}\left(TV\right) $ & $\bar{3}$ & 1 & $ +\frac{2}{3} $ & $ 0 $ 
& $ +\frac{1}{3} $ \\ 
\cline{1-6}
$\bar{\tilde{U}}\left(\bar{T}\bar{V}\right)$ & $3$ & 1 & $ -\frac{2}{3} $ 
& $ 0 $ & $ -\frac{1}{3} $  \\
\hline
\end{tabular}  
\caption{Quantum numbers of dirishonic bound states with $S=1$ and $M=0$.}
\label{tab:dipreons}
\end{center}
\end{table} 
Because dirishons with spin $S=1$ are in a fully 
symmetric representation
in spin space, they must be in a fully antisymmetric representation 
in metaspin space corresponding to $M=0$ to form completely antisymmetric
spin-metaspin inner product states 
\begin{equation}
\Yvcentermath1
\yng(1,1)_{\ SM} = \yng(2)_{\ S} \times \yng(1,1)_{\ M}.  
\end{equation}
The total wave function of dirishons with $S=1$ is then
fully antisymmetric
\begin{equation}
\Yvcentermath1
\yng(1,1)_{\ OHCSM} = \yng(2)_{\ O} \times \yng(2)_{\ HC} 
\times \yng(1,1)_{\ SM},  
\end{equation}
as in the case of fermionic rishon bound states.
 
\subsection{Six-rishon bound states with spin ${\bf S=1}$}
Previously~\cite{abms09}, we have seen that there are transitions
between dirishons and six-rishon bound states with $S=1$. For example, in
hyperbaryon decay, three neutralons $N(VV)$ can combine to 
a $\chi$-boson
\be
3 N \left(VV\right) \longrightarrow 
\chi \left(\begin{array}{c} VVV \\ VVV\end{array}\right).
\ee
Since neutralons have metaspin $M=0$, it follows  
that the $\chi$-boson has metaspin $M=0$ as well. 
Furthermore, as a result of neutrino-antineutrino oscillations~\cite{abms09},
the $\chi$ can oscillate into the weak $Z$-boson and the ${\bar{\chi}}$
\begin{equation}
\label{chiz}
\chi \longleftrightarrow Z \longleftrightarrow \bar{\chi}. 
\end{equation}
Because the $\chi$, $Z$, and other six-rishons with spin $S=1$, 
as well as the dirishons in Table~\ref{tab:dipreons}  
belong to the same supermultiplet of the effective SU(9) gauge 
group~\cite{abms09}, we conclude that all spin 1 gauge bosons,
in particular photons, gluons, and hypergluons 
have metaspin $M=0$. This has consequences for the stability 
of the scalar dirishons to be discussed in the next section.

\section{NWW theorem and the lightest dirishon}
\label{sec:nww}

The Nussinov-Weingarten-Witten theorem~\cite{nww83}, which is 
valid in a QCD like theory, provides an inequality between 
baryon and meson masses. It has been proven using n-flavor lattice 
QCD techniques by Weingarten and independently by Nussinov using gluon 
exchange dynamics and the variational principle. In QCD, the theorem states
that there is a colorless pseudoscalar quark-antiquark bound state 
which is lighter than the lightest colorless three-quark bound state.
More specifically, this leads to the following inequality~\cite{nww83}
\be
m_{\pi} \le \frac{2}{3} m_{N},
\ee
where $m_{\pi}$ and $m_{N}$ are the pseudoscalar pion and nucleon masses.

In QCD these inequalities appear to be connected with spontaneous chiral symmetry breaking S$\chi$SB.
However, in the present vectorlike theory, chiral symmetry is preserved as 
required by 't Hooft anomaly conditions. 
As shown in his original paper, 't Hooft's anomaly matching condition~\cite{gth79}
can be satisfied in a vectorlike theory with only 2 fundamental fermion flavors but not in QCD
with 3 quark flavors. Therefore, in QCD with three (and more) quark flavors,
chiral symmetry is necessarily spontaneously broken. This need not be
the case in the present rishon model based on only 2 elementary (T and V rishon) flavors.
In this model, the presence of the SU(3)$_C$ group
and the possibility of obtaining a third class of fermionic bound states (hyperquarks)
affect the anomaly matching conditions and change the pattern of chiral symmetry
breaking compared to the one of a single hypercolor SU(3)$_H$ group.
Previously we have shown
that for the global U(1) preon number and global B-L symmetries,
't Hooft anomaly matching may be satisfied if a new group of SU(3)$_C$ color singlet
fermionic bound states, called hyperquarks, is introduced~\cite{abms05}.
Thus, there is some evidence that 't Hooft's anomaly matching conditions
are not necessarily in conflict with vectorlike confining theories.
It is therefore not unreasonable to expect
that spontaneous chiral symmetry breaking can be avoided in the present model.

Below we argue that S$\chi$SB is not a necessary condition for the validity of the NWW theorem.
First, the NWW mass inequality $M_B \ge (3/2) \, M_M$  is also valid for
the zero mass case for both mesons and baryons thereby excluding both the
Nambu-Goldstone (massless pseudoscalars, massive fermions) and
the Wigner-Weyl (massive pseudoscalars, massless fermions) modes of
chiral symmetry breaking. Second, baryon-meson mass inequalties can be derived using
a simple one-gluon exchange potential model and the variational priniciple.
In particular, in Nussinov's one-gluon exchange model derivation
the relation $V_{qq} =2/3 V_{q \bar{q}}$ solely based on the evaluation of
the color generators  $\lambda_1\cdot \lambda_2$ for mesons and baryons is essential.
This makes it clear that the validity of meson-baryon mass
inequalities is not limited to theories which display S$\chi$SB.
The latter is a sufficient but not a necessary condition for the validity of the NWW theorem.
This allows the existence of nearly massless pseudoscalar mesons which are not 
the Goldstone bosons connected with S$\chi$SB and of nearly massless fermionic preonic 
bound states.  In Ref.~\cite{geo89} it is clearly stated that S$\chi$SB
implies the existence of massless pseudoscalar particles
but that the converse is not true, i.e. the existence of nearly massless pseudoscalar particles
(e.g. the dirishonic $\zeta$) does not imply that chiral symmetry is spontaneously broken.
As a result, the NWW theorem is valid in theories that do not display spontaneously
broken chiral symmetry.

When applied to rishons, the theorem requires the existence 
of a hypercolor neutral, dirishonic, pseudoscalar bound state with a mass 
smaller than the lightest hypercolor neutral three-rishon bound state 
of the theory, i.e. the neutrino.
The simplest hypercolor neutral, rishon-antirishon bound state
is the pseudoscalar combination composed of 
$T{\bar T}$ and $V{\bar V}$  
\begin{equation}
\zeta = \frac{1}{\sqrt{2}} \left(T\bar{T} - V\bar{V}\right) 
\mbox{.} \nonumber
\end{equation}
We call this particle scalarino $\zeta$. 
There is also an orthogonal state, called scalaron $\sigma$, 
\begin{equation}
\sigma = \frac{1}{\sqrt{2}} \left(T\bar{T} + V\bar{V}\right) \mbox{.} 
\end{equation}
The scalarino $\zeta$ corresponds to the neutral pion 
$\vert \pi^{0}\rangle=(1/\sqrt{2})\,\vert u\bar{u} - d\bar{d}\rangle$,  
and the scalaron $\sigma$ to the heavy scalar meson 
$\vert f^{0} \rangle=(1/\sqrt{2})\,\vert u\bar{u} + d\bar{d}\rangle$ 
in QCD.

As color and hypercolor singlets, the $\zeta$ and 
$\sigma$ have fully symmetric hypercolor-color 
product wave functions. 
Because these spin 0 states are composed of distinguishable particles 
(rishons and antirishons) we cannot infer their metaspin quantum number 
from the requirement of antisymmetry of the total wave function. 
However, we have seen in Sec.~\ref{sec:pauli} 
that dirishons with $S=1$, e.g., the $N(VV)$, must have metaspin $M=0$ 
for the spin-metaspin wave function to be completely antisymmetric. 
Likewise, for fermionic rishon bound states
an antisymmetric spin-metaspin wave function was required.
Extending this pattern to the scalar $\zeta$ and $\sigma$, 
these states must have $M=1$ (see Table~\ref{tab:scalar dirishons}). 
This assignment is supported by considering 
the charged scalar partners of the $\sigma$, which necessarily have
$M=1$, and their metaspin conserving decays in Eq.(\ref{decay}).

Based on the NWW theorem, metaspin conservation, and 
the preon-triality rule, we deduce that scalarinos $\zeta$ 
have the following properties. 

First, for the scalarino mass, the NWW theorem~\cite{nww83} implies 
\be
\label{NWW}
m_{\zeta} \leq \frac{2}{3} \, m_{\nu_{e}}.
\ee
From our estimate for the neutrino mass $ m_{\nu_e} 
\cong 10^{-8} $ eV~\cite{abms09} follows that $m_{\zeta}\leq 10^{-8}$ eV.

Second, because of metaspin conservation, the $\zeta$ with metaspin $M=1$ 
cannot decay into massless photons and gluons because the latter have 
metaspin $M=0$ as discussed in Sec.~\ref{sec:pauli}.  

Third, scalarinos do not interact with fermions below the GUT scale $M_G$.
Otherwise, the small $\zeta$ mass would entail a long-range scalar interaction 
between fermions that is not observed. 
The suppression of fermion-scalarino interactions is guaranteed by the preon 
triality rule~\cite{abms09}. 
The latter states that below energies of order $M_{G}$ vacuum creation and 
annihilation of dirishonic bosons is forbidden; 
only a simultaneous vacuum creation and annihilation of 3 preon-antipreon 
pairs is allowed. Therefore, for $E \leq M_{G}=10^{16}$ GeV one has 
the following constraint (preon triality rule) 
\begin{equation} 
\label{modulo3rule}       
n_{(\bar{T}T)} + n_{(\bar{V}V)} = 3 k 
\end{equation}
where $n_{(\bar{T}T)}$, $n_{(\bar{V}V)} $, and $k$ are natural numbers.
Thus, scalarinos have only been created above the GUT scale 
from rishon-antirishon annihilation into a heavy scalaron and its 
subsequent decay as depicted in 
Fig.~\ref{figure:scalar_interaction},
thereby violating Eq.(\ref{modulo3rule}). 
For the same reason there is no $\zeta \, \bar{\zeta}$ pair annihilation into
two photons or production of $\zeta \,\bar{\zeta}$ pairs from two photons
below the GUT scale. 

The $\zeta$ mass can also be estimated from the relation~\cite{abo83} 
\be 
\label{zeta_mass}
m_{\zeta} = \frac{1}{f_{G}} \ m_{\pi} f_{\pi},  
\ee
where $f_{\pi}\cong 93$ MeV is the pion decay constant and 
$m_{\pi}\cong 140$ MeV is the pion mass. Here, $f_{G}$ is the scale 
where a single rishon-antirishon pair can fuse via a preon-triality
process into a dirishonic 
scalar $\sigma$ as in Fig.~\ref{figure:scalar_interaction}.
Originally, Eq.(\ref{zeta_mass})
with $m_{\zeta}$ and $f_G$ replaced respectively by the axion mass $m_{A}$ 
and the Peccei-Quinn symmetry breaking scale $v$ was derived 
using current algebra techniques to obtain bounds on the mass and 
couplings of the invisible axion. Here, we obtain with 
$f_{G} \geq M_{G} = 10^{16}$ GeV a scalarino mass of 
$m_{\zeta} \leq 10^{-9}$ eV in agreement with the bound obtained 
from the NWW theorem, Eq.(\ref{NWW}). 
The mass of the dirishonic $\sigma$ boson must then be in the range  
$f_{G} <  m_{\sigma} \leq 10^{17}$ GeV. Above $10^{18}$ GeV
no bound state can exist and rishons appear as asymptotically free 
particles~\cite{abms09}.

In summary, the scalarino is the lightest rishon bound state, 
which is furthermore  absolutely stable and inert.

\section{Heavy dirishonic scalars} 
\label{sec:scalar}
In this section we briefly comment on another interesting aspect 
of heavy dirishonic scalar bound states. 
The neutral scalaron $\sigma$ with metaspin $M=1$ 
provides an additional binding force between rishons with metaspin 1/2  
at a mass scale of $m_{\sigma} \cong 10^{17}$ GeV 
described by the Lagrangian
\begin{equation}
\label{scal}
{\cal L}_{\sigma} = g_{M} \, \bar{\psi} \, 
\mbox{\boldmath$\sigma$}_{M} \cdot \mbox{\boldmath$\phi$}_{\sigma}\, \psi, 
\end{equation}
where $g_{M}$ is the rishon-scalaron coupling constant, 
$\b{\sigma}_{M}$ is the metaspin Pauli matrix, 
$\mbox{\boldmath$\phi$}_{\sigma}$ is
the metaspin vector scalaron, and $\psi$ the rishon field.
  
The existence of a short-ranged attractive $\sigma$ exchange interaction 
explains why the dirishonic bosons 
$N(VV)$, $U(T{\bar V})$, and ${\tilde U}(TV)$ with $S=1$ 
and $M=0$ (see Table~\ref{tab:dipreons}) 
are bound with a strength $g_{M} > g_{G}$, where $g_{G}=\sqrt{4\pi \alpha_G}$ 
is the coupling strength of grand unification~\cite{abms09}. 
The binding of these colored and/or hypercolored gauge bosons 
cannot be fully explained by color and hypercolor $SU(3)$ forces and requires
an additional short-range scalar attraction. Without this scalar binding 
the situation would be as in QCD where color unsaturated $qq$ states
do not exist. The same scalar binding mechanism explains the relative
stability of six-rishon bound states, 
e.g. the weak gauge bosons $W$ and $Z$ of the standard 
model and the gauge bosons
of the $SU(6)_{P}$ and $SU(9)_{G}$ gauge groups~\cite{abms09} 
with saturated metaspin $M=0$. 

The $\sigma$-binding force is also responsible for the formation of
dirishonic $\sigma^{+}\left(TT\right)$
and $\sigma^{-}\left(\bar{T}\bar{T}\right)$ bound states 
(see Table~\ref{tab:scalar dirishons}).
Because these states do not occur in the $SU(9)_{G}$ gauge group,
they must be spin scalars, and therefore have metaspin $M=1$ as 
discussed in Sec.~\ref{sec:nww}.
In analogy to fermionic bound states, where $m_e > m_{\nu}$, 
the charged scalars have a larger mass than their neutral counterparts
$m_{\sigma^{\pm}} \geq m_{\sigma^{0}}$. Furthermore, because they participate 
in both the $\sigma$-interaction and in color-hypercolor interactions they are 
expected to have a larger decay width compared to the $\sigma^{0}$.
Their prevalent metaspin conserving decay modes are 
\begin{eqnarray}
\label{decay}
\sigma^{+} &\longrightarrow& \sigma^{0} + U + \tilde{U} \nonumber \\
\sigma^{-} &\longrightarrow& \sigma^{0} + \bar{U} + \bar{\tilde{U}}  \mbox{.} 
\end{eqnarray}
Here, metaspin $M=1$ is carried by the scalars $\sigma^{\pm}$ 
and $\sigma^0$, whereas the vector bosons $U$ and ${\tilde U}$ 
have $M=0$ (see table~\ref{tab:dipreons}).
This completes our discussion of the spectrum of dirishonic scalars.
\begin{table}[htb]
\begin{center}
\begin{tabular}{l c c c c c c}
\hline
Dirishon & H & C & ${\cal{P}}$ & $\Upsilon$ & $Q$ & $ \Pi $ \\ 
\hline
$\sigma^{+}\left(TT\right)$ & $ \bar{3} $ & $ \bar{3} $ & 
$ +\frac{2}{3} $ & $+\frac{2}{3}$ &$+\frac{2}{3}$ & $+1$ \\ 
\hline
$\sigma^{-}\left(\bar{T}\bar{T}\right)$ & $ 3 $ & $ 3 $ & 
$ -\frac{2}{3} $ & $ -\frac{2}{3} $ &$ -\frac{2}{3}$ & $+1$ \\   
\hline
$\sigma^{0}\left(T\bar{T} + V\bar{V}\right)/\sqrt{2} 
$ &$ 1 $ & $ 1 $ & $ 0 $ & $ 0 $ & 
$ 0 $ & $+1$ \\ 
\hline
$\zeta \left(T\bar{T} - V\bar{V}\right)/\sqrt{2} 
$ &$ 1 $ & $ 1 $ & $ 0 $ & $ 0 $ & 
$ 0 $ & $-1$ \\ 
\hline
\end{tabular}  
\caption{Quantum numbers of scalar dirishonic bound states 
with $S=0$ and $M=1$.}
\label{tab:scalar dirishons}
\end{center}
\end{table} 
\section{Metaspin and a compactified fourth spatial dimension}
\label{sec:meta}

With the help of metaspin, we could construct fully antisymmetric 
rishon bound states, and together with the NWW theorem predict
the existence of the lightest dirishonic bound state $\zeta$
but have avoided the question of the origin of this quantum number.
At present we can only speculate that metaspin appears to be connected 
with an extended space-time symmetry involving a compactified fourth spatial 
dimension, as suggested by the isomorphism between 
SU(2)$_S\times$SU(2)$_M\sim$ SO(4), rather than being an internal symmetry. 

This is reminiscent of Kaluza-Klein theory~\cite{kal21}, 
which provides a common framework
for gravitational and electromagnetic forces by extending the number of 
spatial dimensions to four. Kaluza-Klein theory supplies a length 
scale $l$ for the rolled-up fourth spatial dimension~\cite{kle26}
\be
l = \frac{\hbar \, \sqrt{16 \pi G}}{e c} 
= l_{Pl} \, \frac{2}{\sqrt{\alpha_{Q}}} \sim 10^{-33} \ \mbox{m}, 
\ee
where $G$ is Newton's gravitational constant, 
$l_{Pl} = \sqrt{\hbar G/c^{3}}$ 
is the Planck scale of quantum gravity and
$\alpha_{Q} = e^{2}/(4 \pi \hbar c)$ 
the Sommerfeld constant.  The length scale $l$ corresponds to
an energy scale $E \sim 10^{17}$ GeV. 

Interestingly, this is close to the
preon scale $M_{pr} \sim 10^{18}$ GeV, discussed in Ref.~\cite{abms09}. 
It is also in the energy range $10^{17}-10^{18}$ GeV where metaspin plays an 
important role in preonic interactions, in particular 
in the formation of rishonic bound states satisfying the Pauli
principle. At the same energy scale, the metaspin conserving decays of the
charged heavy scalars $\sigma^{\pm}$ into the SU(9) 
gauge bosons $U$ and ${\tilde U}$ and the scalaron
$\sigma^0$ as in Eq.(\ref{decay}) occur. Furthermore,  
at around $10^{17}$ GeV the latter decays into the light scalarinos as shown 
in Fig.~\ref{figure:scalar_interaction}.
Therefore, the length scale where metaspin plays an important role 
coincides with the scale $l$ of the compactified fourth dimension
according to Kaluza-Klein theory. This suggests that the SU(2) metaspin  
symmetry may be connected with the periodicity 
generated by rolling up the fourth spatial dimension.

After leptons, quarks, and hyperquarks
have been formed, metaspin is still a conserved quantum number,  
but ceases to play an active role at larger length scales, e.g., 
in the antisymmetrization of three-quark bound states. Similarly, 
we need not antisymmetrize the quarks in one hydrogen atom with 
those in the other when construcing the wave function 
of the $H_2$ molecule.

\section{Dirishonic cold dark matter}
\label{sec:dark}

The properties of the light pseudoscalar $\zeta$ discussed 
in Sec.~\ref{sec:nww} would make this particle an interesting 
cold dark matter candidate as outlined below. 
Because of the preon triality rule in Eq.(\ref{modulo3rule}), 
the $\zeta$  was produced only primordially at energies 
above $M_{G}$ via the decay of a heavy $\sigma$ boson with mass 
$m_{\sigma} > M_{G}$ as shown in Fig.~\ref{figure:scalar_interaction}. 

The large mass of the $\sigma$  
implies that scalarinos were created before quarks and leptons were 
formed.  On the other hand, due to their small mass, scalarinos 
may have been more 
copiously produced than any other rishonic bound state in the early Universe. 
An inflatory expansion near the GUT scale decelerated 
the scalarinos to velocities close to the nonrelativistic limit.
Furthermore, the $\zeta$ has metaspin $M=1$ and thus cannot decay 
into massless photons and gluons with $M=0$ due to metaspin conservation.
As a result, below the GUT scale, the $\zeta$ appears as an absolutely 
stable, inert, nearly massless particle.

What additional cosmological insights can be drawn from this scenario?
Given the scalarino mass deduced from Eq.(\ref{NWW}) and 
Eq.(\ref{zeta_mass}), we can estimate the number of scalarinos 
in the Universe following the arguments in Ref.~\cite{nus85}. 
It is known~\cite{wmap11} that the ratio of baryonic density  
to the dark matter density of the Universe is approximately 
$\rho_{B}/\rho_{DM}\cong 0.2$. Assuming that scalarions are the only source
of dark matter, i.e., $\rho_{DM}=\rho_{\zeta}$ we get 
for the corresponding matter densities the following expressions 
\bea 
\rho_{B} &=& n_{B} \, m_{B} = \epsilon_{B} \,  n_{\gamma}\, m_{B} \nonumber \\
\rho_{\zeta} &=& n_{\zeta} \, m_{\zeta}, 
\eea
where $n_{B}$, $n_{\gamma}$, and $n_{\zeta}$ are the number densities 
of baryons, photons, and scalarinos. 
Furthermore, $m_{B} \cong 1$ GeV is the baryon mass 
and $\epsilon_{B}=n_{B}/n_{\gamma}= 6 \,\, 10^{-10}$ is the baryon 
asymmetry~\cite{bar03}.  
We then get with $m_{\zeta} \cong 10^{-9}$ eV 
\be
\frac{n_{\zeta}}{n_{\gamma}}= 
\frac{\rho_{\zeta}}{\rho_{B}} \, \, 
 \frac{m_{B}}{m_{\zeta}}\, \, \epsilon_{B} = 3 \, \, 10^{9}.
\ee
Therefore, the ratio of scalarino to 
photon number density is approximately equal to the ratio of photon to 
baryon number density, i.e., we have $10^{9} $ times more scalarinos 
than photons per volume. This suggests a very   
homogeneous distribution of dirishonic cold dark matter in the Universe.

\section{Summary}
\label{sec:sum}
To satisfy the requirement of total antisymmetry of rishon bound state
wave functions we have introduced an SU(2) like 
quantum number $M$, called metaspin. We could then 
construct totally antisymmetric fermionic and bosonic rishon bound states. 
In each case, the antisymmetry resides in the spin-metaspin product space. 
As a result, $S= 1$ bosons necessarily have 
$M=0$, whereas $S=0$ bosons have $M=1$. 
Due to the isomorphism SU(2)$_S$ $\times$ SU(2)$_M\sim$ SO(4), 
it has been suggested that metaspin could be connected with the 
existence of a fourth spatial dimension, which is compactified and 
limited to distances near the Planck scale.

Based on the Nussinov-Weingarten-Witten theorem applied to
the rishon model, we have deduced the existence of a  pseudoscalar 
dirishonic meson, called scalarino $\zeta$, 
that is lighter than the lightest neutrino and hence is the lightest 
rishon bound state. We have estimated the scalarino mass 
to be of order $m_{\zeta} \cong 10^{-9}$ eV.

The stability, abundance, and inertness 
of scalarinos has been derived from the following arguments.
First, due to $M$ spin conservation, scalarinos with $M=1$ 
cannot decay into massless gauge bosons with $M=0$. 
Second, scalarinos were created 
only at energies above the GUT scale of $M_G\cong 10^{16}$ GeV 
with a production rate that vastly exceeds that of any other rishon 
bound state. 
We have estimated the ratio of cosmic scalarino and photon number densities 
as $n_{\zeta}/n_{\gamma} \sim 10^9$. 
Third, below the GUT scale, scalarinos interact with other particles 
only via gravitational interaction and induced higher order electromagnetic 
interactions, which will make their detection difficult. 
Nevertheless, it appears that its properties make the scalarino 
a viable cold dark matter candidate.

\acknowledgments{ We thank H. Harari,  
D. Lichtenberg, and \hfill \break I. Obukhovsky for useful discussions.}


\begin{thebibliography}{9}
\bibitem{comment1} To satisfy the 't Hooft anomaly matching conditions,
a new group of fermionic rishon bound states, 
called hyperquarks, was introduced~\cite{abms05}. This led to the condition  
$N_G=N_C=N_H$, i.e., the equality of the number of generations $N_G$, 
colors $N_{C}$, and hypercolors $N_H$.  Hyperquarks were later shown to 
play an important role in explaining the left-right asymmetry of 
low-energy weak interactions as a residual force on the rishon bound state 
level~\cite{abms09}.
\end{thebibliography}
\end{document}